\documentclass[12pt,a4paper]{article}
\usepackage[latin9]{inputenc}
\usepackage{color}
\definecolor{note_fontcolor}{rgb}{0.800781, 0.800781, 0.800781}
\usepackage{amsmath}
\usepackage{graphicx}
\usepackage{authblk}

\makeatletter

\providecommand{\tabularnewline}{\\}
\newenvironment{lyxgreyedout}
  {\textcolor{note_fontcolor}\bgroup\ignorespaces}
  {\ignorespacesafterend\egroup}
\usepackage{tikz}
\usetikzlibrary{calc}

\usepackage{fullpage}
\usepackage{sidecap}
\usepackage{cite}

\makeatother

\begin{document}

\title{An efficient multi-scale Green's Functions Reaction Dynamics scheme }

\author[1]{Luigi Sbailò}
\author [1]{Frank Noé }
\affil[1]{Department of Mathematics and Computer Science, Freie Universit\"at
Berlin, Arnimallee 6, 14195 Berlin, Germany}
\maketitle
\begin{abstract}
Molecular Dynamics - Green's Functions Reaction Dynamics (MD-GFRD)
is a multiscale simulation method for particle dynamics or particle-based
reaction-diffusion dynamics that is suited for systems involving low
particle densities. Particles in a low-density region are just diffusing
and not interacting. In this case one can avoid the costly integration
of microscopic equations of motion, such as molecular dynamics (MD),
and instead turn to an event-based scheme in which the times to the
next particle interaction and the new particle positions at that time
can be sampled. At high (local) concentrations, however, e.g. when
particles are interacting in a nontrivial way, particle positions
must still be updated with small time steps of the microscopic dynamical
equations. The efficiency of a multi-scale simulation that uses these
two schemes largely depends on the coupling between them and the decisions
when to switch between the two scales. Here we present an efficient
scheme for multi-scale MD-GFRD simulations. It has been shown that
MD-GFRD schemes are more efficient than brute-force molecular dynamics
simulations up to a molar concentration of $10^{2}\mu M$. In this
paper, we show that the choice of the propagation domains has a relevant
impact on the computational performance. Domains are constructed using
a local optimization of their sizes and a minimal domain size is proposed.
The algorithm is shown to be more efficient than brute-force Brownian
dynamics simulations up to a molar concentration of $10^{3}\mu M$
and is up to an order of magnitude more efficient compared with previous
MD-GFRD schemes.
\end{abstract}

\section{Introduction}

Particle-based reaction-diffusion simulations have been widely used
to simulate signaling cascades in biological systems \textbf{\cite{SchoenebergEtAl_BJ14_PhototransductionKinetics,eGFRD,ErbanChapman_PhysBiol09_StochasticReactionModelling,GFRD2,SchoenebergEtAl_NatComm17_SNX9}}.
In contrast to other approaches to simulate molecular kinetics simulations,
such concentration-based approaches or Gillespie's dynamics \textbf{\cite{Gillespie_JCP76_Gillespie,ST-CME}},
the trajectory of all interacting particles is resolved, providing
a reaction kinetics model with high spatio-temporal detail. Particles
diffuse according to the Langevin equation, and whenever they are
close to each other reactions can happen. In a brute-force approach,
all particles are simultaneously propagated over a fixed integration
step \textendash{} at sufficiently long timescales typically using
a time-discretization of the overdamped Langevin, or Brownian dynamics
(BD) equation \cite{langevin}. Unfortunately, a short integration
step is generally required to avoid systematically missing particle
interactions \textbf{\cite{smoldyn1}}. In interacting-particle reaction-diffusion
(iPRD) simulations, particles are interacting with nonlinear potentials
at close distances, which requires even shorter time steps in the
BD integrator \textbf{\cite{SchoenebergNoe_PlosOne13_ReaDDy,SchoenebergUllrichNoe_BMC14_RDReview}}.
Especially in biological applications, where many proteins may interact
in a crowded environment to give rise to some supramolecular machinery,
such detailed simulations may be required \cite{GunkelEtAl_Structure15_Rhodopsin,UllrichEtAl_PLoSCB11_ActiveZone,McGuffeeElcock_PlosCB10_Crowding}.
However, this approach becomes computationally expensive with large
particle numbers, and also when fast-diffusing species are involved
that require small simulation time steps, making it challenging to
reach biologically relevant time scales \textbf{\cite{SchoenebergEtAl_BJ14_PhototransductionKinetics,BiedermannEtAl_BJ15_ReaddyMM}}.
Hence, designing efficient multi-scale reaction-diffusion algorithms
that can reach the biologically relevant resolution where required,
but avoid unnecessary computation time wherever possible, is of high
relevance for the bio-simulation community.

One possible strategy adopted to improve computational performances
in particle-based simulations is implementing an event-based algorithm
as in the first-passage kinetic Monte Carlo (FPKMC) algorithm \cite{FPKMC,FPKMC2,FPKMC3}
and Green's functions reaction dynamics (GFRD/eGFRD) \cite{GFRD1,GFRD2,eGFRD}.
The central idea is to   directly sample the next time point at which
particles will interact, e.g. to perform a reaction, rather than simulating
the trivial diffusion of free particles via BD. GFRD  is synchronous
and approximate: in every iteration of the algorithm, an integration
step length is chosen such that at most two particles can interact;
particles are propagated for that time and eventually react \cite{GFRD1,GFRD2}.
Depending on the system configuration, a new integration step is selected.
This algorithm may suffer from inaccuracies because a finite propagation
time always results in a finite choice of interactions between more
than two particles simultaneously, which is not covered by the algorithm.

In the subsequent asynchronous versions, firstly proposed in FPKMC
\cite{FPKMC,FPKMC2,FPKMC3} and then in eGFRD \cite{eGFRD}, the volume
of the system is decomposed into non-overlapping protective domains
containing one or at most two particles. In each of these domains
a next event is sampled. Events comprise domain escapes, unimolecular
reactions, or bimolecular reactions in domains containing two particles.
In this asynchronous scheme, a list of all scheduled events is initially
compiled, then at every step the system jumps to the next event and
the list gets updated with a new event. However, some unscheduled
events can occur and the list must then be updated on the fly. For
example, when a particle is about to enter a protective domain, this
domain must be burst, i.e. destroyed, the particle positions must
be sampled prematurely, and new protective domains must be drawn.

A recent extension of this algorithm is the multi-scale combination
of explicit time-step integration (for the sake of generality called
molecular dynamics (MD), although in many practical cases BD will
be used) and FPKMC/eGFRD, in short MD-GFRD \cite{MD-GFRD,GFRDanis}.
In MD-GFRD, interacting particles, i.e. particles that are close in
space, are simulated via short time steps, whereas isolated particles
are propagated via an event-based FPKMC/eGFRD scheme on longer time
scales, protective domains thus can contain only one particle. Using
direct time-integration at short distances allows to incorporate a
variety of effects that are relevant to describe molecular detail.
For example, these local dynamics could involve momenta \cite{MD-GFRD},
anisotropic diffusion \cite{SchluttigEtAl_PRE10_AnisotropyProteinProtein,GFRDanis},
nonlinear interaction potentials or complex reactions \cite{SchoenebergNoe_PlosOne13_ReaDDy},
and would be a natural place to include the dynamics simulated by
kinetic models obtained from all-atom MD, e.g. Markov State Models
(MSMs) \cite{PrinzEtAl_JCP10_MSM1,BowmanPandeNoe_MSMBook,SarichSchuette_MSMBook13,NoeClementi_JCTC15_KineticMap,PlattnerEtAl_NatChem17_BarBar}
or multi-ensemble Markov models (MEMMs) \cite{WuMeyRostaNoe_JCP14_dTRAM,WuEtAL_PNAS16_TRAM}. 

MD-GFRD has been shown to be several order of magnitudes faster than
brute-force integration of Brownian dynamics \cite{MD-GFRD,GFRDanis}.
The efficiency improvement is particularly evident in dilute systems,
where particles spend most of their time freely diffusing in the system
before encountering each other, which renders an event-based algorithm,
that directly samples encountering times, dramatically faster. However,
this efficiency is lost at high densities, while the efficiency of
direct time-step integration is only mildly dependent on the particle
density (e.g. through the number of neighbor interactions that need
to be evaluated in each time step). Indeed, constructing a domain
and sampling an event in it is computationally more demanding than
performing few brute-force Brownian motion steps. Therefore, one typically
avoids the construction of very small domains that would burst rapidly,
and instead uses direct time-step integration when the size of a newly
constructed domain is below the minimal domain size \cite{MD-GFRD,GFRDanis,Sokolowski}.
Still, as the system becomes more dense, the efficiency of this scheme
decreases, as the fraction of particles that are described by direct
time-step integration increases, and domains, which are required to
be non-overlapping, tend to be smaller and thus more prone to a premature
burst. In this context, determining the optimal size of the minimal
domain and avoiding unnecessary, premature bursts can be critical
to ensure computational performance.

In this paper, we present a domain making scheme and several numerical
improvements that make multi-scale FPKMC/eGFRD algorithms such as
MD-GFRD more efficient. The main developments are the determination
of the optimal domain size upon construction and of the minimal domain
size for the construction of small domains. 

\section{Molecular Dynamics - Green's Function Reaction Dynamics}

We briefly introduce into MD-GFRD in order to summarize the concepts
relevant for the present paper. In MD-GFRD, the system is decoupled
into non-overlapping spherical domains, or shells, that contain at
most one particle. MD-GFRD is an event-based algorithm, whose events
are particle escapes from their protective domain. The event times
are obtained by sampling from a Green's function as explained below. 

Brownian motion can be described probabilistically by the Einstein
diffusion equation,

\begin{equation}
\frac{\partial p(\vec{r},t)}{\partial t}=D\thinspace\Delta p(\vec{r},t),\label{eq:diff}
\end{equation}
where $p(\vec{r},t)$ is the probability distribution of a Brownian
particle with diffusion coefficient $D$, $\vec{r}=(r,\,\theta,\,\phi)^{\top}$
is the position of the particle and $\Delta$ is the Laplace operator
in spherical coordinates. Isolated particles are treated using Green's
function dynamics. To facilitate that, one creates spherical ``protective''
domains of radius $b$ around them, in order to mark the volume within
which they can diffuse without interacting with other particles. The
domain size $b$ is chosen such that it contains only one particle
and the whole sphere's volume is not subject to any external potentials,
i.e. the interaction of other particles, membranes, etc.  Given the
spherical symmetry of this problem, the evolution of the probability
distribution can be described by the radial function $p(r,t)$, which
represents the probability to be in any point on the surface of a
sphere of radius $r$. The radial probability to be at a radius $r<b$,
without having previously hit the domain border $b$, is computed
by imposing absorbing boundary conditions on the domain borders, $p(b,t)=0$
\cite{Redner}. By imposing this boundary condition and the initial
condition $p(r_{0},t_{0})=\delta(r_{0})$ on Eq. \eqref{eq:diff},
we obtain:

\begin{equation}
p(r,t|r_{0}=0,t_{0})=\frac{1}{S(t)}\sum_{m=1}^{\infty}\exp{\bigg\{-m^{2}\frac{\pi^{2}D}{b^{2}}(t-t_{0})\bigg\}}\frac{2\pi r}{b^{2}}\thinspace m\thinspace\sin\bigg(\frac{m\pi r}{b}\bigg),\qquad r<b.\label{eq:p(r,t)}
\end{equation}
$S(t)$ is the survival probability

\begin{equation}
S(t|r_{0}=0,t_{0})=-2\sum_{n=1}^{\infty}(-1)^{n}\thinspace\exp{\bigg\{-n^{2}\frac{\pi^{2}D}{b^{2}}(t-t_{0})}\bigg\},\label{eq:S(t)}
\end{equation}
which represents the probability that the particle is inside the domain
at $t$, without having previously hit the borders. The first exit
time probability $q(t)$ is defined \emph{via} the survival probability
$S(t)$

\begin{equation}
q(\tau|r_{0}=0,t_{0})=-\frac{dS(\tau)}{d\tau}=-2\sum_{n=1}^{\infty}(-1)^{n}\exp{\bigg\{-n^{2}\frac{\pi^{2}D}{b^{2}}(\tau-t_{0})}\bigg\}\frac{n^{2}\pi^{2}}{b^{2}}D,\label{eq:q(t)}
\end{equation}
and it gives the probability that the particle escapes its domain
for the first time at $\tau$.

In this derivation, we have assumed that no other particles enter
the domain, and the particle inside the domain is not subject to any
external potentials or forces (e.g. exerted by particles near the
domain). However, in a multi-particle simulation this assumption is
not always valid. Let us assume that at $t_{0}$ we have constructed
a protective domain around an isolated particle, and this particle
has sampled a first exit time $t_{0}+\tau$ from its domain. In this
situation, it is possible that an external particle, whose motion
is brute-force integrated, is in proximity to the first domain at
a time, $t_{1}<t_{0}+\tau$, \emph{i.e.} before the escape time. The
first exit time $\tau$ has been sampled assuming that no other particle
interact with the domain, hence the intrusion of another particle
before that time would make the sampling of the particle's escape
time invalid. Consequently, to ensure that particles in protective
domains are freely diffusing, we define a burst radius for each pair
of particles to be at least the interaction length between the intruding
particle and the particle in the domain.  Whenever a particle approaches
a protective domain to a distance below the burst radius the domain
is burst, i.e. destroyed. In that event, the particle position is
updated inside the domain by sampling eq. \eqref{eq:p(r,t)} at time
$t=t_{1}$. After a domain burst, the clock of the two particles is
synchronized to $t_{1}$.

\subsection{Algorithm outline}

\begin{figure}
\centering{}\centering \includegraphics{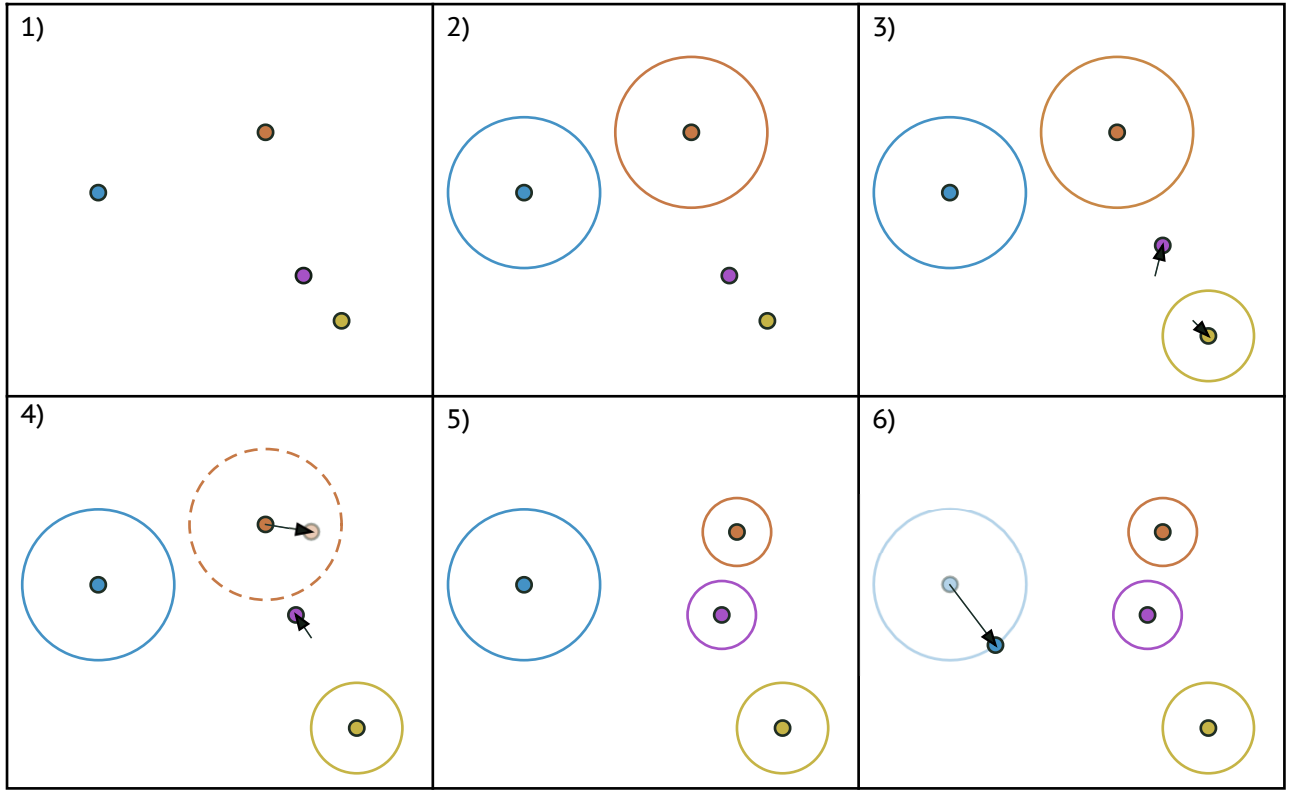} \caption{Outline of the multi-scale MD-GFRD algorithm. 1) Particles are placed
in their initial configuration. 2) A protective domain is drawn on
all those particles that are not directly interacting.  Domains are
effectively constructed only if their size is larger than the particle's
minimal domain size (blue and orange particles). 3) Particles which
don't have a protective domain are integrated with direct time steps
(purple and yellow particles), and as soon as a particle becomes sufficiently
distant from all others, a protective domain is generated around it
(yellow particle).  4) When a particle gets too close to a domain
(purple particle), this is burst and the inside particle samples a
new position (orange particle). 5) After a domain burst, the new
particle position can be sufficiently far from the intruding particle
to allow both particles for constructing a protective domain (orange
and purple particles). 6) The global time advances to the next time
from the scheduled exit times, and the exiting particle position is
sampled randomly from all points on the previous protective domain
(blue particle).}
\label{Fig:outline} 
\end{figure}

In MD-GFRD, the particle propagation is performed alternatively via
direct time-step integrations or Green's functions samplings. The
choice of the propagation method depends on the system configuration
and, in particular, whether the particle is freely diffusing or interacting
with other particles. At each iteration of the algorithm, one particle
is selected from a time-ordered event-list. If this particle is not
interacting with other particles, the construction of a protective
domain is attempted. The construction is then accepted only if the
domain radius is larger than the minimal domain size, whenever the
construction is rejected the particle motion is instead brute-force
integrated. 

In this scheme, particle interactions are always evaluated on discrete
times $\left\{ t_{n}\right\} $, where $t_{n}=n\thinspace dt$, $n$
is an integer, and $dt$ is the MD integration step. Therefore, a
GFRD particle that leaves a protective domain and thus becomes an
MD particle is mapped to the next discrete time via a small Brownian
motion step. MD particles that are evaluated at the same time point
$t$ can be updated simultaneously and collectively as in usual MD
implementations. In the following pseudocode, however, it is simpler
to explain the algorithm as if all particles are treated by an asynchronous
event list. 

Each particle possesses a \emph{current} and a \emph{scheduled} position
and time. Each particle is also associated with an event, that takes
the particle from its current position and time to its scheduled position
and time, if it is successfully executed. Events include MD integration
step and scheduled exits from a protective domain, but they may be
modified due to events such as domain bursting. In the beginning
of the simulation, the domain making algorithm creates a protective
domain for each particle that is not involved in a direct interaction.
Domains larger than the minimal domain size $\rho$ are constructed,
and first exit times are sampled via Eq. \eqref{eq:q(t)}. These exit
events are then stored in a list ordered by increasing scheduled-time.
All particles that could not construct a protective domain are placed
on top of the event-list, forces between them are computed and their
scheduled positions are computed and stored. Based on this initial
list, the following asynchronous algorithm propagates the system state
in time:
\begin{enumerate}
\item Pick the first particle $i$ in the event-list: 
\begin{enumerate}
\item If the particle was in a protective domain: place it on a position
sampled uniformly at random on the domain boundary. Then, propagate
it to the next discrete time $t_{i}$ via a free Brownian motion sampling.
\item Else: update the particle position and time to the stored scheduled
position and time.
\end{enumerate}
\item Compute the distances $\left\{ r_{ij}\right\} _{j=1}^{N}$ from the
$N$ neighboring particles. The distances are between the centers
of mass and are computed between synchronous positions when particles
are not located in a protective domain; otherwise, the distance between
the center of mass of the particle $i$ and the center of the protective
domain of the particle $j$ is computed. 
\item For all $j=1,...,N$:  if the particle $j$ is in a protective domain
and the $i-j$ distance is below the burst radius ($r_{ij}-r_{j}<R_{burst}^{i}$,
where $r_{j}$ is the domain size of the particle $j$): 
\begin{enumerate}
\item Burst the $j$-domain.
\item Synchronize the scheduled time of particle $j$ to $t_{i}$ and update
the scheduled position of particle $j$ by sampling from Eq. \eqref{eq:p(r,t)}.
\item Place particle $j$ on top of the event-list.
\item Update the $r_{ij}$ distance.
\end{enumerate}
\item Use the distances $\left\{ \widetilde{r}_{ij}\right\} _{j=1}^{N}$
, where $\widetilde{r}_{ij}=r_{ij}-R_{int}^{ij}$ and $R_{int}^{ij}$
is the interaction length, in a domain making algorithm to create
a domain with radius $r_{i}$: 
\begin{enumerate}
\item If the proposed radius is larger than the minimum domain size, $r_{i}>\rho_{i}$:
accept the domain, sample the first exit time $\tau_{i}$ from Eq.
\eqref{eq:q(t)}, and increase the particle event time by $\tau_{i}$.
\item Else: Update the  scheduled position and scheduled time via direct
time-step propagation (this step might involve also interactions and
reactions).
\end{enumerate}
\item Place the particle $i$ in the event-list according to increasing
event time.
\end{enumerate}

Note that if particles $i$ and $j$ construct domains that are in
contact and if following step 1a these particles have identical scheduled
discrete exit times, it is possible that the particle $i$, upon escape,
bursts the $j$-domain at a later time than the scheduled exit time
of particle $j$. This apparent inconsistency is due to the fact that
in this serial algorithm particle $j$ has not executed the step 1a
yet. Clearly, in this occasion the position of particle $j$ is updated
by executing the step 1a rather than sampling from Eq. (\ref{eq:p(r,t)}).

In Fig. \ref{Fig:outline}, a graphical representation of a possible
outcome of this algorithm is shown; there is not a match between the
points in the algorithm and the points in the figure. 

\section{Domain making scheme and minimal domain size}

The basic idea of domain making schemes is that larger domains correlate
with more efficient computation, as the particle doesn't participate
in direct time-step integration during the correspondingly longer
exit times (see Eq. \eqref{eq:q(t)}). However, choosing domain sizes
in a greedy manner does not necessarily lead to optimal performance.
For instance, when a large domain is next to a much smaller one, or
to a domain close to its escape time, the latter domain is likely
to experience a particle exit very soon, which might in turn burst
the large domain, thereby annihilating the advantage of the long exit
time from that domain. Domain bursting is not convenient, since it
involves sampling a second Green's function. Moreover, it represents
an unscheduled event that is difficult to treat efficiently in a parallel
implementation.

The minimal domain size determines whether the domain construction
is accepted or not. Instead of sampling the first exit time from a
small domain, it might be more convenient to simulate the same particle
propagation via direct time-step integrations. Indeed, solving a first
exit time problem has generally a higher computational cost than simulating
a number of direct time-step integrations. Thus, in MD-GFRD algorithms,
the dimension of the smallest domain whose construction is allowed
must be determined: whenever the construction of a domain of smaller
size is attempted, this trial is rejected and the particle is instead
brute-force integrated.

\subsection{MD-GFRD}

\label{subsec:md-gfrd}

\begin{figure}[t]
\centering \includegraphics{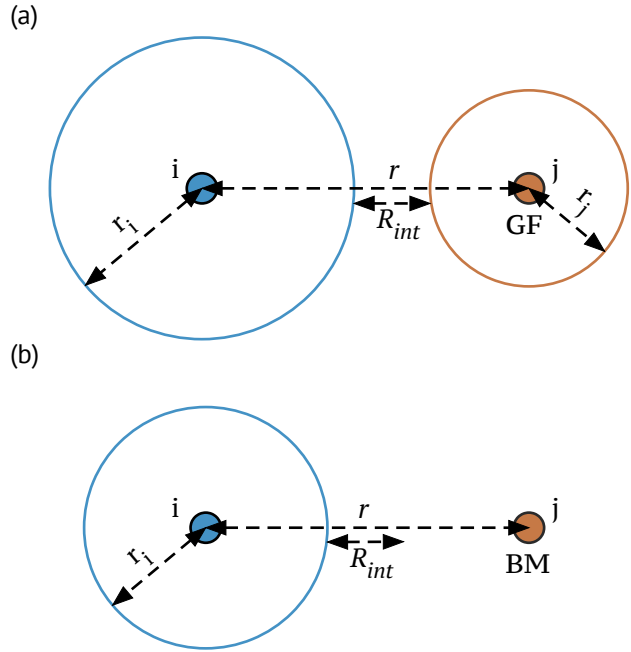} \caption{MD-GFRD domain making scheme suggested in \cite{MD-GFRD,GFRDanis}.
The domain size choice is made according to the status of the neighboring
particle: a) the prime neighbor is a GF particle, then the shell takes
all available space; b) the prime neighbor is a BM particle, then
only half of the available space is used.}
\label{Fig:MD-GFRDsizing} 
\end{figure}

The MD-GFRD domain making schemes employ the largest shell principle
to draw protective domains. We distinguish between Green's function
(GF) particles which are located in a protective domain  and Brownian
motion (BM)  particles that are undergoing a direct time-step integration.
The domain making routine firstly computes the center-center distance
$r_{ij}$ between the particle $i$ of interest from all neighboring
particles $j$, subtracting the interaction length $R_{int}^{ij}$
of the particle pair. The resulting distance $\widetilde{r}_{ij}=r_{ij}-R_{int}^{ij}$
is then divided by 2 if the particle $j$ is a BM particle. If the
particle $j$ is a GF particle, the distance is reduced by the \textit{$j$}-domain
size $r_{j}$ (Fig. \ref{Fig:MD-GFRDsizing}). In the case of a BM
particle only half of the total distance is used to let the other
particle construct a domain of equal size in the subsequent step.
This routine is iterated over all neighboring particles and the lowest
value obtained is finally selected. This domain creation makes domains
as large as possible while avoiding direct particle interaction.

In previous studies, the minimal domain size in MD-GFRD algorithms
has been set proportional to the particle radius \textbf{\cite{MD-GFRD,GFRDanis,Sokolowski}}, where the sum of the particles radii gives the particles pairwise
interaction. In particular, the minimal domain size has been suggested
to be always larger or equal than the particle radius \cite{Sokolowski}.
In the implementation of Ref. \cite{GFRDanis}, the minimal domain
size $\rho$ is chosen to be equal to the particle radius. In the implementation of Ref.
\cite{MD-GFRD}, $\rho$ can have different values depending on whether
the particle is undergoing a direct time-step integration ($\rho_{GFRD}$)
or has just escaped a protective domain ($\rho_{BD}$). The minimal
domain value assumes a larger value when the particle is under direct
time-step integration ($\rho_{GFRD}>\rho_{BD}$). This technique has
been used to prevent particles from rapidly switching between the
GF and BM mode. Indeed, when the particle motion is subject to direct
time-step integration, it is likely to be located in a crowded region
of the system, where a domain is more likely to be burst. Diminishing
the number of domains constructed in this regions correlates with
a lowering of the total number of bursts. This scheme has been used
to simulate particles interacting via a Lennard-Jones potential, and
the minimal domain values $\rho{}_{GFRD}=5\sigma$ and $\rho_{BD}=3\sigma$
were used, where $\sigma$ is the Van-der-Waals radius. 

Finally, the bursting radius  should be chosen equal or larger than
the interaction length of the two particles. However, it cannot be
larger than the minimal domain size of any other particle to prevent
the algorithm from entering in an infinite mutual bursting loop, where
a pair of isolated particles alternatively construct a domain which
is burst by the other particle in the subsequent step. In MD-GFRD,
the bursting radius is set equal to the interaction length plus 
the minimal domain size of the particle, because whenever a particle
is close to another domain, that domain must be burst in order to
allow creating two new domains of significant size.

\subsection{New domain-making scheme\label{subsec:domainMak}}

\begin{figure}[t]
\centering \includegraphics{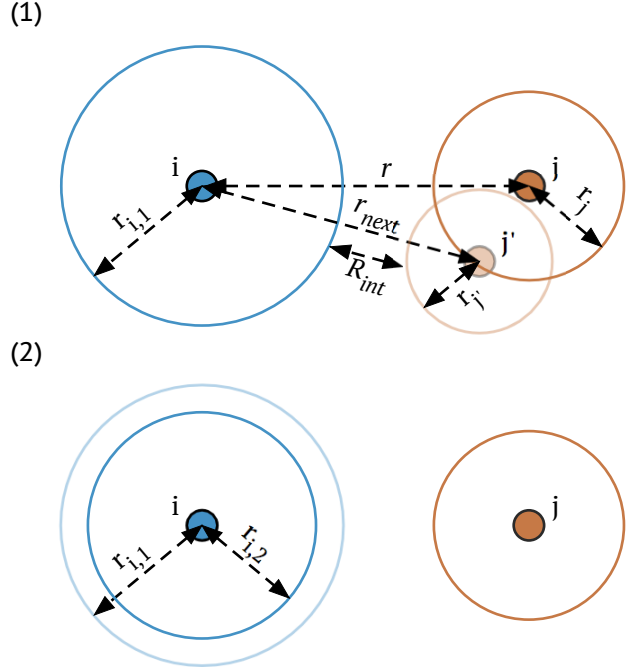} \caption{New domain making scheme for the case of an isolated pair of particles.
At time $t_{i}$, particle $i$ is attempting the construction of
the $i$-domain close to particle $j$ that is already enclosed in
a domain. The escape time $t_{j'}>t_{i}$ and particle $j$'s escape
position were sampled when the domain was constructed. 1) Particle
$i$ constructs a domain whose size $r_{i,1}$ is such that its average
first exit time is the same as the average first exit time of particle
$j$ from the $j'$-domain that might be constructed after the exit
from its current $j$\textendash domain. 2) The domain size in the
previous step obtained is further reduced to finally obtain $r_{i,2}$. }
\label{Fig:NewScheme} 
\end{figure}

The aim of the new scheme is to improve the algorithm's computational
performance and to decrease the number of domain bursting events.
In order to keep the number of bursting events small, domains are
sized such that they have the same average first exit time as the
domains that will be constructed in their proximity. The key idea
is that when domains are constructed, not only the first exit time
of the particle is sampled, but also its exit position. This information
is used by neighboring particles to propose an optimized domain size
such that it has the same average first exit time as the domains that
will be later constructed on the memorized exit positions (Fig. \ref{Fig:NewScheme}
1). In Ref. \cite{FPKMC3} the importance of constructing optimized
domains has already been discussed, and it is suggested that domains
should be constructed to delay in time as far as possible the first
event in the queue, which corresponds to constructing domains with
equal mean first exit times. However, this was achieved only when
all domains are constructed simultaneously, which optimizes only over
the first event in queue. By pre-sampling the exit position of particles,
it is instead possible to construct balanced domains over a long series
of events. Although developed for MD-GFRD, the idea of pre-sampling
the exit position can also be applied to FPKMC/eGFRD schemes. 

In order to further reduce the number of bursting events, the domain
size is then shrunk. Although the domains are not chosen to be of
maximum size, this approach significantly reduces the overall number
of bursts compared to the scheme described in Sec. \ref{subsec:md-gfrd}.
The choice for the size reduction in the second step (Fig. \ref{Fig:NewScheme}
2) is performed to obtain a balance between a low number of bursts
and long domain exit times. Clearly, the specific setting of these
parameters depends on implementation details such as serial or parallel
execution etc, and can be adapted to the local setting.  This algorithm
is illustrated in the simplest case of an isolated pair of particles
in Fig. \ref{Fig:NewScheme}. In the new scheme, the bursting radius
is also chosen to be equal to the interaction length plus the minimal
domain size.

In practice, if the domain is created close to a GF particle (Fig.
\ref{Fig:NewScheme} 1) the first domain $r_{i,1}$ is obtained by
solving a system of two equations:
\begin{align}
\frac{r_{j'}^{2}}{6D_{j}}+\Delta t & =\frac{r_{i,1}^{2}}{6D_{i}},\label{eq:system}\\
\tilde{r}_{\mathrm{next}} & =r_{i,1}+r_{j'},
\end{align}
where $\tilde{r}_{\mathrm{next}}=r_{\mathrm{next}}-R_{\mathrm{int}}$
is the available space, $r_{\mathrm{next}}$ is the distance between
the center of particle $i$ and the exit position of particle $j$,
$\Delta t=t_{j'}-t_{i}$ is the time difference between the scheduled
exit time of particle $j$ and the current time, \emph{i.e.} the time
in which particle $i$ is attempting to construct a domain.  The
first equation imposes that the average exit time from the $i$-domain
is the same as from the $j'$-domain, where the expected exit time
$\langle\tau\rangle$ of a Brownian particle with diffusion coefficient
$D$ from a sphere of radius $b$ is:

\begin{equation}
\langle\tau\rangle=\frac{b^{2}}{6D}.\label{eq:avTime}
\end{equation}
The second equation enforces the domains to be adjacent by taking
all available space, according to the largest shell principle. In
contrast to MD-GFRD, the largest domain principle is applied between
the $i$-domain and the $j'$-domain that is possibly constructed
subsequently.

If the average first exit time of particle $i$ from the available
space $r_{i,1}=\widetilde{r}_{\mathrm{next}}$ is less than $\Delta t$,
the time interval to the scheduled exit time of particle $j$, the
solution of the system in Eq. (\ref{eq:system}) has no real values,
which means that the $i$-domain and the $j'$-domain cannot have
the same average first exit time. As the $j$-particle is not expected
to burst the $i$-domain in this case, we use all available space
for the $i$-domain, i.e. $r_{i,1}=\widetilde{r}_{next}$. Consistently,
inserting $\Delta t=\widetilde{r}_{next}^{2}/6D_{i}$ in Eq. \eqref{eq:system}
results in the solution $r_{i,1}=\widetilde{r}_{next}$.

The system in Eq. (\ref{eq:system}) is then solved only when $\Delta t<\tilde{r}_{\mathrm{next}}^{2}/6D_{i}$.
The optimal domain size is then given by:

\begin{equation}
r_{i,1}=\begin{cases}
\widetilde{r}_{\mathrm{next}}, & \Delta t\ge\frac{\widetilde{r}_{\mathrm{next}}^{2}}{6D_{i}}.\\
\widetilde{r}_{\mathrm{next}}\,\frac{1-\sqrt{1-(1-\frac{D_{j}}{D_{i}})(1+\frac{6\,\Delta t\,D_{j}}{\widetilde{r}_{next}^{2}})}}{1-\frac{D_{j}}{D_{i}}}, & \mathrm{otherwise}.
\end{cases}\label{eq:newGF-2}
\end{equation}
The square root argument in Eq. \eqref{eq:newGF-2} is always positive
if $\Delta t<\widetilde{r}_{next}^{2}/6D_{i}$, therefore the solution
is always real-valued. The boundary condition $0<r_{i,1}<\tilde{r}_{next}$
has been applied, as explained in Appendix \ref{App:boundCond}.

If the two particles have identical diffusion coefficients $D$, the
solution simplifies to:
\begin{equation}
r_{i,1}=\frac{\widetilde{r}_{\mathrm{next}}}{2}+\frac{3\thinspace D\thinspace\Delta t}{\widetilde{r}_{\mathrm{next}}}.
\end{equation}
The value $r_{i,1}$ obtained is a function of the distance $\widetilde{r}_{\mathrm{next}}$.
Hence, $r_{i,1}$ does not take the volume of the existing $j$-domain
into account and thus does not ensure to avoid overlap of the $i$
and $j$ domains. To avoid such an overlap, the $i$-domain must be
accordingly resized to the largest possible value: $r_{i,1}=\widetilde{r}-r_{j}$,
where $\tilde{r}=r-R_{\mathrm{int}}$ and $r$ is the center-center
distance between particles $i$ and $j$.

A similar approach is used if particle $j$ is a BM particle. In this
case, the $i$-domain is created so as to leave enough space for particle
$j$ to construct a domain whose first exit time is equal to the $i$-domain:

\begin{equation}
r_{i,1}=\frac{\widetilde{r}}{1+\sqrt{\frac{D_{j}}{D_{i}}}}.\label{eq:newBM}
\end{equation}
Finally, the domain radius is further reduced as :

\begin{equation}
r_{i,2}=r_{i,1}-n_{\mathrm{red}}\sqrt{2D_{j}\,dt},.\label{eq:newRED}
\end{equation}
where $n_{\mathrm{red}}$ is a parameter (Fig. \ref{Fig:NewScheme}
2). The domain reduction is set proportional to the average displacement
that the particle $j$ performs in one integration step. This reduction
is performed to reduce the probability that the particle $j$ bursts
the $i$-domain in cases where the sampled escape time of the particle
$i$ is larger than the expected value. Note that if $\Delta t>r_{i,1}^{2}/6D_{i}$
the particle $j$ is expected to escape its domain after the particle
$i$, in this case there is no need to reduce the size of the $i$-domain
and thus the step in Eq. (\ref{eq:newRED}) is omitted.  When this
scheme is applied to multi-particle systems, the previously outlined
approach is applied to all nearest-neighbor particle pairs, and the
lowest value of $r_{i,2}$ is chosen.

\subsection{New scheme for minimal domain size\label{subsec:minDomain}}

\begin{figure}
\centering \includegraphics[width=8.5cm]{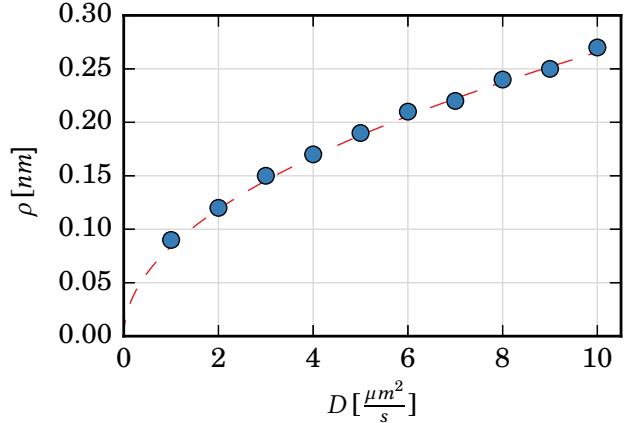} \caption{Minimal domain radius $\rho$ as a function of $D$ using the time
step $\mathrm{d}t=0.1\thinspace ns$. The dots represent the radius
of the minimal protective domain where Green's functions sampling
and direct time-step integration have equal CPU costs. Simulations
to compute the first exit time from the domain with size $\rho$ were
conducted for different diffusion coefficients and domain sizes, using
either direct time-step integration or Green's functions sampling.
For small domain sizes, the direct time-step integration is always
more efficient. The dashed red line shows $\rho=\alpha\thinspace\sqrt{D\thinspace dt}$,
as described in Eq. \eqref{eq:rho}, with the implementation-specific
value $\alpha=8.4$ that has been found empirically. }
\label{Fig.GF-BM} 
\end{figure}

In contrast to previous works, the minimal domain size is proposed
here to be proportional to the square root of the particle diffusivity,
rather than the particle size. The minimal domain size defines the
particle distance below which direct time-step integration is assumed
to be more efficient than sampling Green's functions. We assume that
the CPU time required to sample the probability density of the first
exit time is approximately independent of domain size and diffusion
coefficient. In contrast, the CPU time spent to simulate first exit
times via brute-force integrations depends on the domain size, on
the particle diffusion coefficient and on the time-step length.

Given the average first exit time $\langle\tau\rangle$ of a particle
with diffusion coefficient $D$ from a sphere of radius $b$, Eq.
\eqref{eq:avTime}, the average number of steps $\langle n\rangle$
to simulate the first exit time is:
\begin{equation}
\langle n\rangle=\frac{b^{2}}{6\thinspace D\thinspace dt},\label{eq:<n>}
\end{equation}
where $dt$ is the time step. The average CPU time, $\langle T_{BF}(b)\rangle$,
spent to compute escape times via brute-force integrations is proportional
to the number of integration steps, and thus:
\begin{equation}
\langle T_{BF}(b)\rangle\propto\frac{b^{2}}{D\thinspace dt}.
\end{equation}
It is assumed that the average CPU time, $\langle T_{GF}(b)\rangle$,
spent to sample a Green's function is approximately constant.
\begin{equation}
\langle T_{GF}(b)\rangle=Const.
\end{equation}
Let $\rho$ be the domain size at which the CPU times are equal, $\langle T_{BF}(\rho)\rangle=\langle T_{GF}(\rho)\rangle$,
then:
\begin{equation}
\rho^{2}\propto D\thinspace dt.
\end{equation}
Hence, the minimal domain radius $\rho(D,dt)$ is defined as the threshold
that determines whether the domain construction is accepted or not.
\begin{equation}
\rho(D,dt)=\alpha\sqrt{D\thinspace dt}.\label{eq:rho}
\end{equation}
Simulations indicate that this function correctly describes the point
where a direct time-step integration becomes more efficient than a
Green's function root finding (Fig. \ref{Fig.GF-BM}). The parameter
$\alpha$ is a value that depends on the implementation and machine,
and is determined in the beginning of a simulation (see Appendix \ref{App:alpha}).

\section{Results}

We compare the performance of the multi-scale MD-GFRD scheme implemented
in Refs. \cite{MD-GFRD} and \cite{GFRDanis}, the new scheme, and
a direct time-step integration scheme using Brownian dynamics. Two
versions of the new scheme are simulated, one with $n_{red}=5$ in
Eq. (\ref{eq:newRED}) (new scheme 1), and one which does not use
domain size reduction ($n_{red}=0$, new scheme 2), thus tending to
size domains more greedily. In addition, we also test a hybrid scheme,
which implements the minimal domain size as described in Sec. \ref{subsec:minDomain}
but employs the same domain making scheme as proposed in Refs. \cite{MD-GFRD}
and \cite{GFRDanis}. For simplicity we simulate  particles in a periodic
box and interacting with a harmonic repulsion:
\begin{equation}
V(r)=\frac{1}{2}\thinspace k\thinspace(R_{int}-r)^{2},\qquad r<R_{int,}\label{eq:potential-1}
\end{equation}
where $r$ is the inter-particle distance between the centers of mass,
$k=100$ is the spring constant, and the interaction length $R_{int}$
is equal to the sum of particle radii.  Reactions, more complex particle-particle
potentials, or other near-space interactions can be straightforwardly
integrated in the direct time-step integration regime that is used
to simulate interacting particles. 

Two simulations have been performed using different diffusion coefficients
and particle radii: 
\begin{enumerate}
\item 10 spherical particles with radius $R=2.5\thinspace nm$ and diffusion
coefficient $D=10\,\mu m^{2}/s$. 
\item 5 faster and smaller particles with radius $R_{1}=1.5\,nm$ and diffusion
coefficient $D_{1}=10\thinspace\mu m^{2}/\mathrm{s}$ and 5 slower
and larger particles with radius $R_{2}=3.5\,nm$ and diffusion coefficient
$D_{2}=1\thinspace\mu m^{2}/\mathrm{s}$.
\end{enumerate}

\subsection{Efficiency comparisons of different MD-GFRD schemes and direct Brownian
dynamics}

To obtain clean benchmarks, most calculations are run with ten particles
and direct evaluation of all pairwise particle distances, while the
particle density is adjusted by choosing the box size. For a more
complex test, Sec. \ref{subsec:neighborlist} simulates larger particle
numbers with a neighbor list implementation.

\begin{figure}
\centering \includegraphics{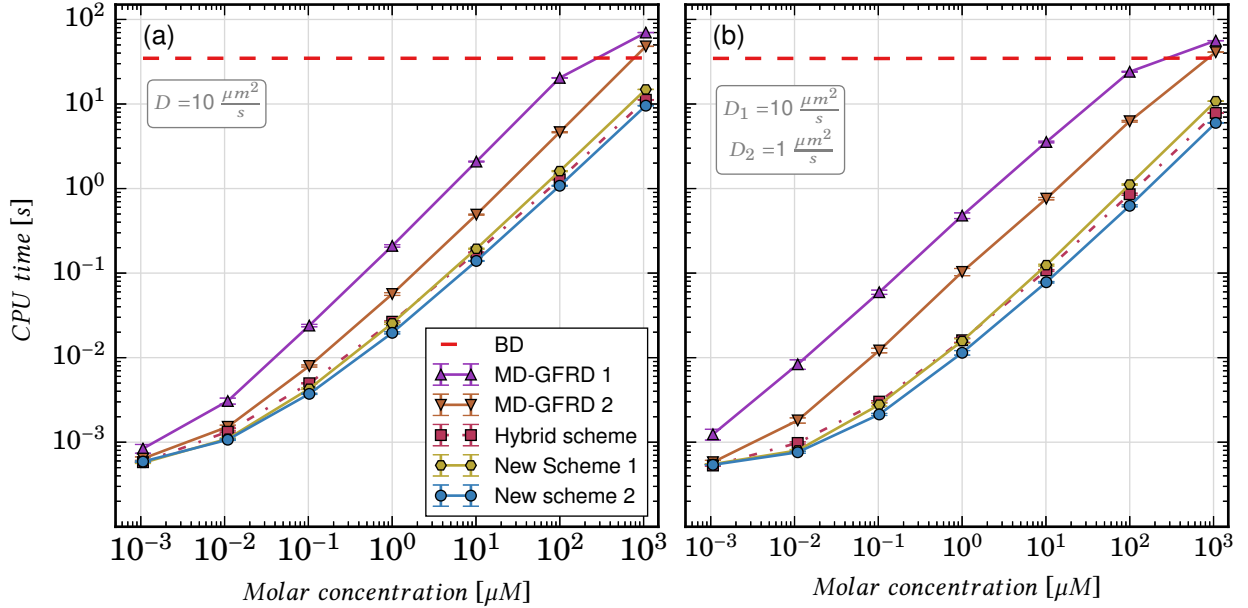} \caption{CPU time required to simulate $1\thinspace ms$  of real time, using
a brute-force integration step of $dt=0.1\thinspace ns$. The number
of particles is kept fixed to $N=10$, while the system volume is
adapted to the selected molar concentration. Simulations are performed
in a cubic-shaped box with periodic boundary conditions. Particles
are spherical-shaped with radius $R=2.5\thinspace nm$ and diffusion
coefficient $D=10\,\mu m^{2}/s$ in (a) and radii $R_{1}=1.5\,nm$
and $R_{2}=3.5\,nm$ and diffusion coefficients $D_{1}=10\,\mu m^{2}/s$
and $D_{2}=1\,\mu m^{2}/s$ in (b). A binary interaction length is
defined as the sum of particles radii, when particles are in within
this distance repulse according to a harmonic potential as in Eq.
(\ref{eq:potential-1}), where $k=100$. The minimal domain of the
new schemes and of the hybrid scheme uses the pre-factor $\alpha=9$
as defined in Eq. (\ref{eq:rho}), see Appendix \ref{App:alpha}.
In new scheme 1, $n_{red}=5$; in new scheme 2, $n_{red}=0$, see
Eq. (\ref{eq:newRED}). In MD-GFRD 1, the minimal domain size is equal
to the particle radius\cite{GFRDanis}. In MF-GFRD 2, the minimal
domain sizes $\rho_{GFRD}=2.5\,R$ and $\rho_{BD}=1.5\,R$\cite{MD-GFRD}
were used, where the pre-factors $1.5$ and $2.5$ have been chosen
to adapt to a different simulation the pre-factors used in Ref. \cite{MD-GFRD},
while preserving their same relative proportions. At low concentrations,
MD-GFRD schemes are several order of magnitude faster than BD. The
new schemes and the hybrid scheme are faster than BD up to concentrations
of $10^{3}\thinspace\mu M$, while MD-GFRD schemes are preferable
over BD up to $10^{2}\thinspace\mu M.$ }
\label{Fig:perf-1} 
\end{figure}

The efficiency of MD-GFRD strongly depends on the particle concentration,
since in case of dilute systems particles are allowed for constructing
large domains and performing large time steps. Hence, MD-GFRD algorithms
are dramatically faster than BD schemes at low concentrations. As
the particle concentration is increased, MD-GFRD becomes less efficient,
while the BD efficiency remains constant. Consequently, there is a
concentration threshold where BD starts being more efficient than
MD-GFRD. In Fig. \ref{Fig:perf-1}, the performance is compared between
the new schemes, the hybrid scheme, the previous MD-GFRD schemes and
direct BD simulation. It is evident that all MD-GFRD schemes are several
order of magnitude faster than BD at low densities. Moreover, the
new schemes are faster than the previous MD-GFRD schemes at all densities,
but performances are similar at low densities. In particular, for
both diffusion coefficients, the new schemes and the hybrid scheme
are preferable over BD for concentrations up to $10^{3}\:\mu\mathrm{M}$,
whereas previous MD-GFRD schemes were preferable over BD only up to
molar concentrations of $10^{2}\:\mu\mathrm{M}$. The schemes which
implement the new minimal domain size all show similar performance,
and among them the new scheme 2 is the fastest. We note that these
numbers may be different in different implementations (codes and machines),
and comparison is therefore only meaningful within the same implementation.

\begin{figure}
\centering \includegraphics{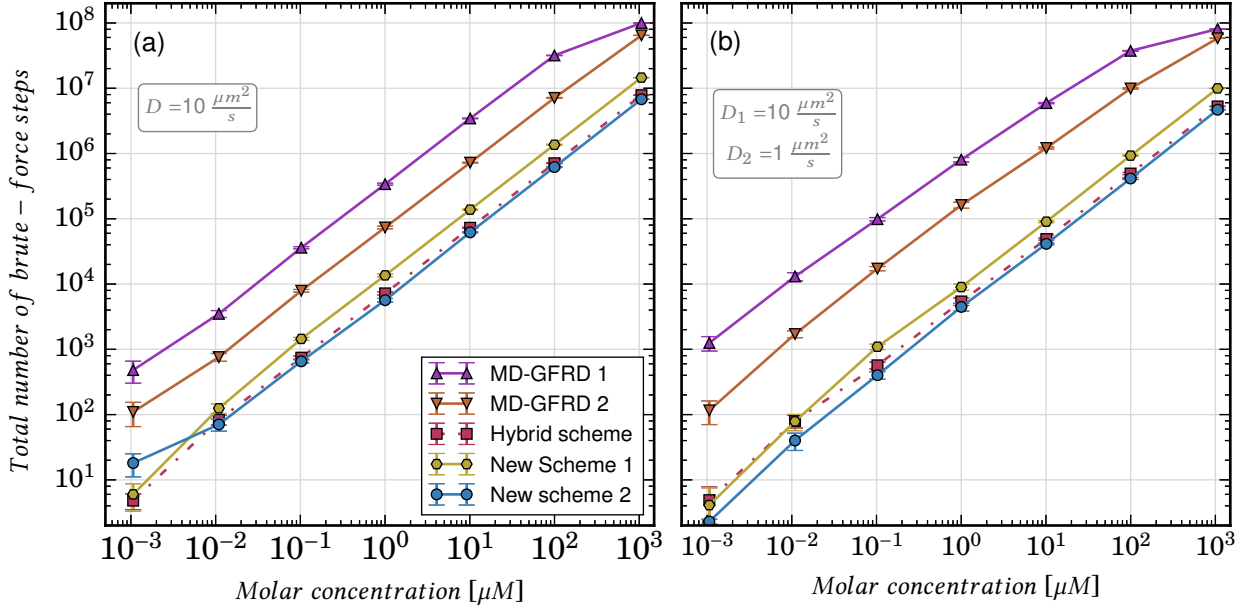} \caption{Total number of direct time-steps in the multi-scale MD-GFRD simulations
described in Fig. \ref{Fig:perf-1}. As the particle density increases,
interactions between particles become more frequent, and more simulation
time is spent in conducting direct time-step integration. The behavior
of these curves is similar to that in Fig. \ref{Fig:perf-1}, indicating
that the number of brute-force Brownian motion steps represent the
bottle-neck in the present simulations. The largest value in each
plot, $10^{8}$, represents the condition where each of the $10$
particles have performed $10^{7}$ direct time-steps, which means
that no particle propagation was made using Green's functions sampling.
In the new schemes and in the hybrid scheme, FPKMC/eGFRD steps are
still done 90$\%$ of the time under the same conditions. }
\label{Fig:BM-1} 
\end{figure}

The total number of direct integration time-steps performed in each
multi-scale MD-GFRD simulation increases with increasing particle
concentration (Fig. \ref{Fig:BM-1}). This growth is remarkably similar
to the growth in the CPU time, indicating that the reason of the improved
performance of MD-GFRD schemes is essentially due to a reduction of
the direct time-integration steps that represent the computational
bottleneck. In the new schemes and in the hybrid scheme, the minimal
domain size is smaller than in previous MD-GFRD schemes, which enables
more protective domains to be constructed, which in turn reduces the
fraction of time spent in direct time-step integrations. Although
having equal minimal domain size, the new scheme 2 shows a slightly
lower number of direct integration time-steps with respect to the
hybrid scheme. This is essentially the result of the construction
of more balanced domains which allow for an optimization of the available
space. On the other hand, the new scheme 1 spends a larger fraction
of time under direct time-step integration, because after the reduction
step more domains are not sufficiently large for construction. 

\newpage{}

\subsection{Minimization of the domain burst frequency}

\begin{figure}
\centering \includegraphics{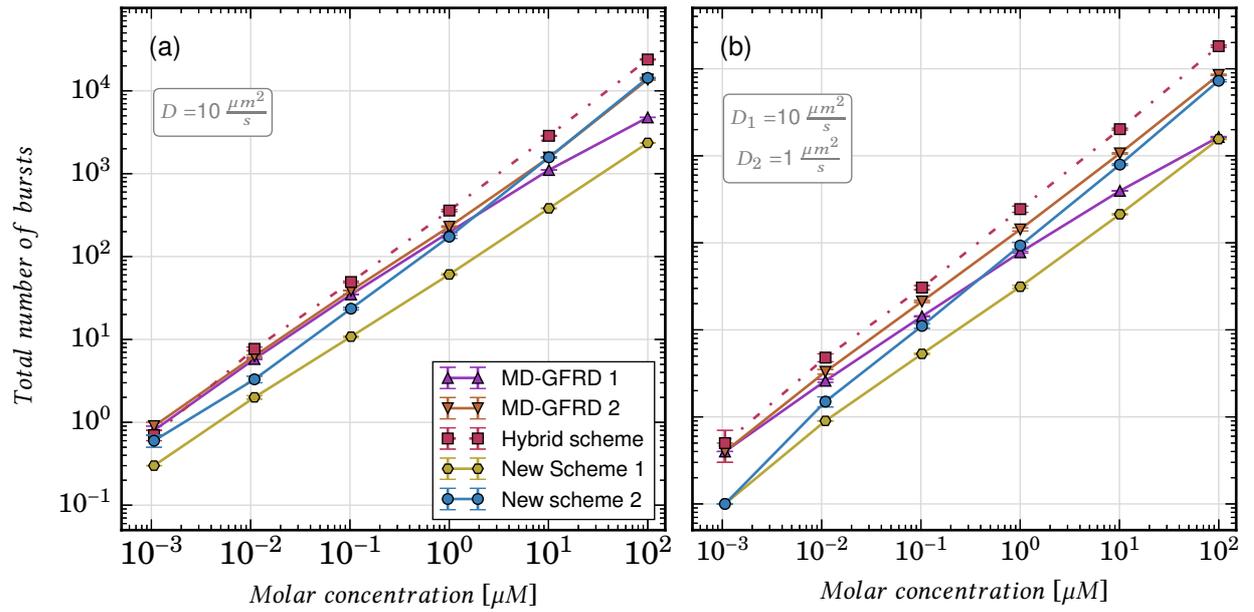} \caption{Average total number of protective domain bursts in the multi-scale
MD-GFRD simulations described in Fig. \ref{Fig:perf-1}. As the molar
concentration is increased, domains tend to be smaller and to be constructed
more often, which goes along with an increase of the number of bursts.
The average number of bursts in the new scheme 1 is lower than in
the previous MD-GFRD implementations at any density. Keeping the
total number of bursts low can be important for efficient parallelization,
e.g. using Graphical Processing Units (GPUs).}
\label{Fig:burst} 
\end{figure}

\begin{figure}
\centering \includegraphics{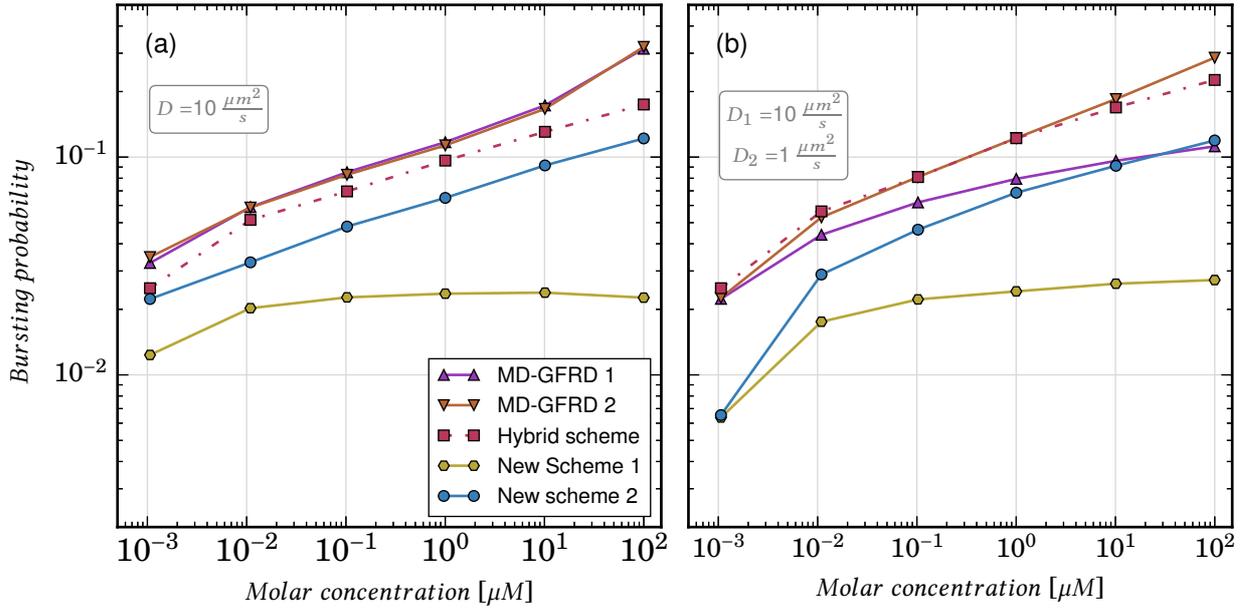} \caption{Domain bursting probability, i.e. ratio of the total number of domains
burst over the total number of domains constructed in the MD-GFRD
simulations described in Fig. \ref{Fig:perf-1}. The domain construction
schemes proposed here are clearly more efficient than previous schemes
and results in domains that are more likely to survive until the particles
contained therein make successful exits. The  bursting probability
is always lower than $3\%$ in the new scheme 1, while in the implementations
MD-GFRD 1,2 this value is roughly an order of magnitude larger at
the higher concentrations. }

\label{Fig:Eff} 
\end{figure}

 Despite the fact that domain sizes are small on average, Fig. \ref{Fig:burst}
shows that the total number of bursts is the lowest in new scheme
1, i.e. when the domain reduction is included. The hybrid scheme involved
the highest number of bursts, since the construction of small domains
is allowed, but their sizes are not chosen optimally. The incorporation
of particle exit positions into domain construction, and  the choice
of domain sizes so as to balance the exit times allows to reduce the
number of bursts to one third (new scheme 2); if a reduction step
is also added (new scheme 1), the number of bursts is further reduced
by approximately one order of magnitude. This improved efficiency
on the domain construction is evident in Fig. \ref{Fig:Eff}, which
shows the probability that a protective domain is burst prematurely
by intrusion of another particle rather than being annihilated by
a regular exit of the particle contained therein. This quantity is
computed as the ratio of the total number of domain bursts over the
total number of constructed domains. At low concentrations the bursting
probability is small, but it increases with increasing particle density.
The new domain-making scheme clearly results in more efficient domains
that are much less probably to be burst prematurely compared to the
previous MD-GFRD scheme, especially at higher concentrations.

The full implementation of the new scheme (version 1) is to be preferred
to previous MD-GFRD schemes in both cases: when the serial computational
performance is most relevant and when the number of total bursts is
required to be low. The MD-GFRD implemented in Ref. \cite{GFRDanis}
is faster than the implementation in Ref. \cite{MD-GFRD}, while the
latter scheme has a lower number of domain bursts. The new scheme
1 is instead superior in both computational performance and number
of domain bursts. More specifically, the implementation as in new
scheme 1 is optimal to drastically lower the number of bursts while
preserving efficiency. The new scheme 2 instead has a slightly higher
CPU performance in our implementation, but does not keep the number
of bursts small. The improvements result to up an order of magnitude
of gain in the CPU performance and an order of magnitude of gain in
the total number of bursts. 

\newpage{}

\subsection{Large particle numbers}

\label{subsec:neighborlist}

\begin{lyxgreyedout}
\end{lyxgreyedout}

The general trends observed in the benchmarks shown in the previous
sections are also expected to hold for systems with many particles.
However, in systems with many particles $n$, it is necessary to implement
a neighbor list to avoid that each timestep scales with $n^{2}$ as
a result of the pairwise distance calculations.

In order to validate that our MD-GFRD scheme can still be efficiently
implemented with many particles, we implemented new scheme 1 with
$n_{red}=5$ using a neighbor list. Particles are interacting with
harmonic repulsion with radius $R=2.5\thinspace nm$, $k=100$, and
periodic boundary conditions are applied as described in the previous
section. The system volume is kept fixed to $17.576\thinspace10^{6}\thinspace nm^{3}$,
while the number of particles is adapted to achieve the desired molar
concentration. All particles have diffusion coefficient $D=10\thinspace\mu m^{2}/s$. 

In order to efficiently implement a neighbor list, we used a discretization
of the simulation box in cells of length $L_{cell}=5\thinspace nm$
for the brute-force BD simulations and of $L_{cell}=10\thinspace nm$
for the MD-GFRD simulations. Each particle checks the cell it is located
in and the 26 neighboring cells for possible neighbors. In such a
cell discretization, the smallest distance at which two particles
can loose track of each other is the cell length, and thus the maximum
protective domain size must be limited to at most half the cell length
minus the interaction length, which is the gap to be left between
contiguous domains. Here, we limited the maximum domain size to $R_{max}=2.5\thinspace nm$.

The simulation results in Tab. \ref{Tab:neighb} show that the new
scheme remains to be faster than a brute-force integration up to a
molar concentration of $10^{3}\mu M$

\begin{table}
\caption{Computational time to simulate $1\thinspace ms$ of real time, using
a brute-force integration step of $dt=0.1\thinspace ns$. Particles
are spherical-shaped with radius $r=2.5\thinspace nm$ and diffusion
coefficient $D=10\,\mu m^{2}/s$. A harmonic potential is used to
reproduce particles interactions as in Eq. (\ref{eq:potential-1}),
with $k=100$. A linked list cell has been implemented, where each
grid box is a cube with length $L_{box}=5\thinspace nm$ in BD, and
$L_{box}=10\thinspace nm$ in the new scheme. In these simulations
the new scheme remains to be more efficient than BD up to a molar
concentration of $10^{3}\mu M$.}
\label{Tab:neighb}

\begin{tabular}{|c|c|c|c|}
\hline 
Molar concentration & Particles number & CPU time, new scheme & CPU time, BD\tabularnewline
\hline 
\hline 
$10^{2}\thinspace\mu M$ & $10^{3}$ & $271\thinspace s$ & $14.5\thinspace10^{3}\thinspace s$\tabularnewline
\hline 
$10^{3}\thinspace\mu M$ & $10^{4}$ & $230\thinspace10^{3}s$ & $260\thinspace10^{3}s$\tabularnewline
\hline 
\end{tabular}
\end{table}

\newpage{}

\subsection{Mean square displacement}

\begin{figure}
\centering\includegraphics{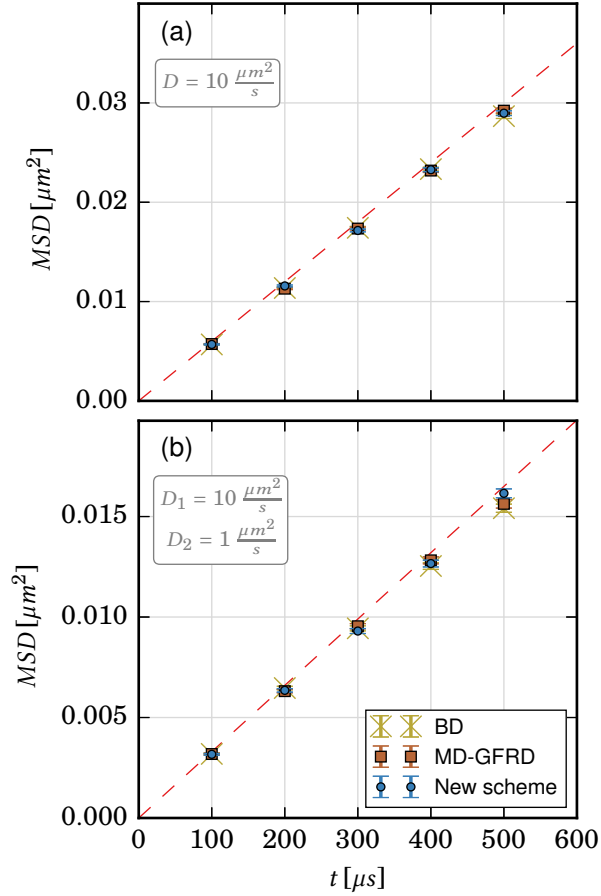}

\caption{Mean square displacement of particles diffusing, simulated as described
in Fig. \ref{Fig:perf-1} for a molar concentration of $10^{2}\mu M$.
The MD-GFRD scheme used is from Ref. \cite{GFRDanis}, in the new
scheme a reduction step was performed with $n_{red}=5$. The expected
value of free diffusing particles (red dashed line) is given by $\langle\Delta r^{2}\rangle=6Dt$
in (a) and $\langle\Delta r^{2}\rangle=6\,\frac{D_{1}+D_{2}}{2}\,t$
in (b). The mean squared displacements are slightly below the mean
square displacements of freely diffusing particles due to crowding
effects.}

\label{Fig:Diff}
\end{figure}

In order to validate the implementation of the MD-GFRD schemes, of
the new scheme and of the direct time-step integration scheme used,
the mean squared displacement of the particles simulated with the
different schemes has been recorded and compared. In Fig. \ref{Fig:Diff},
the mean square displacement shows an excellent agreement  between
the different schemes.

\newpage{}

\section{Conclusions}

We have described a novel multi-scale MD-GFRD scheme to simulate diffusion
and interaction of Brownian particles. In a multi-scale MD-GFRD scheme,
the propagation of free particles is performed in an event-based fashion
via Green's functions samplings, whilst the reactions and the interactions
between particles are simulated via direct time-step integration (here
using time-discretized Brownian dynamics, BD).

Multi-scale MD-GFRD has been shown to be several orders of magnitude
faster than BD at low particle concentrations. The efficiency of MD-GFRD
strongly depends on the density of the system, and previous schemes
have been shown to be more efficient than BD up to a molar concentrations
of $10^{2}\mu M$ \cite{MD-GFRD,GFRDanis}. In crowded systems, free
space around particles tends to be scarce and constructing protective
domains around them is more difficult. In addition, domains are often
burst prematurely by the intrusion of other particles, which is undesirable
as it increases the computational effort and the domain making is
less parallelizable than direct BD steps or FPKMC/eGFRD extractions.
It is thus desirable to optimize the domain making scheme so as to
avoid unnecessary premature bursting and improve the computational
performance at a given particle concentration.

In the multi-scale MD-GFRD scheme described in this paper, a new domain
making algorithm and a way to determine the minimal domain size accurately
have been introduced. The new domain making algorithm constructs domains
with sizes chosen so as to balance the domain exit times of adjacent
particles. In contrast to previous domain selection schemes, this
approach involves sampling exit positions, i.e. it looks ahead in
time in order to plan domain sizing optimally. In addition, the minimal
domain size is proposed to be proportional to the square root of the
particle diffusivity, which leads to the existence of smaller domains
than in previous implementations. Nonetheless, the domains created
with this algorithm are more efficient as they are less likely to
burst. Overall, the new scheme exhibits up to an order of magnitude
improvement of computational efficiency compared to the previous
multi-scale MD-GFRD implementations. Moreover, the new scheme is superior
to direct time-step integration for concentrations up to $10^{3}\mu M$.
In future studies, this algorithm will be used as a part of the software
ReaDDy to simulate realistic biological systems.

\subsection*{Acknowledgement}

The authors gratefully acknowledge funding by Deutsche Forschungsgemeinschaft
(SFB 1114/C03 to L.S. and F.N), European Research Commission (starting
grant 307494 ``pcCell'' to F.N.), and the Max Planck Society (International
Max Planck Research School CBSC fellowship to L.S.). The authors would
like to thank Thomas R. Sokolowski for useful discussions.

\appendix

\section{New domain size scheme}

\label{App:boundCond} The solution to Eq. \eqref{eq:system} has
the following two roots:
\begin{equation}
r_{i,1}=\widetilde{r}_{next}\,\frac{1\pm\sqrt{1-(1-\frac{D_{j}}{D_{i}})(1+\frac{6\,\Delta t\,D_{j}}{\widetilde{r}_{next}^{2}})}}{1-\frac{D_{j}}{D_{i}}}.\label{eq:generalSol}
\end{equation}
Assuming that the condition $\Delta t<\tilde{r}_{\mathrm{next}}^{2}/6D_{i}$
is satisfied, the argument of the square root is nonnegative, resulting
in two real-valued solutions. In the following derivations, we study
two different cases depending on $D_{i}$ and $D_{j}$.

Firstly, we study $D_{i}>D_{j}$, which leads to $1-\frac{D_{j}}{D_{i}}>0$.
In case the discriminant is added the factor that multiplies $\tilde{r}_{next}$
is clearly higher than one, since diffusion coefficients are always
positive, then we would obtain $r_{i,1}>\widetilde{r}_{next}$, an
unphysical solution. The discriminant must thus be subtracted. Furthermore,
imposing the condition $\frac{\tilde{r}_{next}^{2}}{6D_{i}}>\Delta\tau$,
or equivalently $\frac{\Delta\tau}{\widetilde{r}_{next}^{2}}<\frac{1}{6D_{i}}$,
we can verify that if the discriminant is subtracted:
\begin{equation}
r_{i,1}<\widetilde{r}_{next}\,\frac{1-\sqrt{1-(1-\frac{D_{j}}{D_{i}})(1+\frac{D_{j}}{D_{i}})}}{1-\frac{D_{j}}{D_{i}}}=\tilde{r}_{next}.
\end{equation}
The condition $r_{i,1}<\tilde{r}_{next}$ is satisfied if the discriminant
is subtracted.

In case $D_{j}>D_{i}$, then $1-\frac{D_{j}}{D_{i}}<0$:
\begin{equation}
\ r_{i,1}=\widetilde{r}_{next}\,\frac{1\mp\sqrt{1+\mid1-\frac{D_{j}}{D_{i}}\mid\,(1+\frac{6\,\Delta t\,D_{j}}{\widetilde{r}_{next}^{2}})}}{\mid1-\frac{D_{j}}{D_{i}}\mid}.
\end{equation}
 In order to satisfy the condition $r_{i,1}>0$, the discriminant
must have a positive sign. However, the sign of the discriminant has
been inverted by the modulus in the denominator, since it comes from
the subtraction of the discriminant.

To sum up, only the root obtained by subtracting the discriminant
satisfies the condition $0<r_{i,1}<\tilde{r}_{next}$:

\begin{equation}
r_{i,1}=\widetilde{r}_{next}\,\frac{1-\sqrt{1-(1-\frac{D_{j}}{D_{i}})(1+\frac{6\,\Delta t\,D_{j}}{\widetilde{r}_{next}^{2}})}}{1-\frac{D_{j}}{D_{i}}}.\label{eq:generalSol}
\end{equation}

\section{$\alpha$ values}

\label{App:alpha}

The minimal domain size is given by eq. (\ref{eq:rho}), where $\alpha$
is a parameter that is determined in the beginning of the simulation.
An optimal value $\alpha=8.4$ has been already suggested in Fig.
\ref{Fig.GF-BM}. However, that value was selected by taking only
the Green's function solver and the direct time-step integrator into
account. In general, it might seem appropriate to insert a penalty
for the possibility of a burst and then to slightly rise the $\alpha$
value, where the penalty would be higher when a higher number of bursts
is expected.

\begin{figure}
\centering \includegraphics{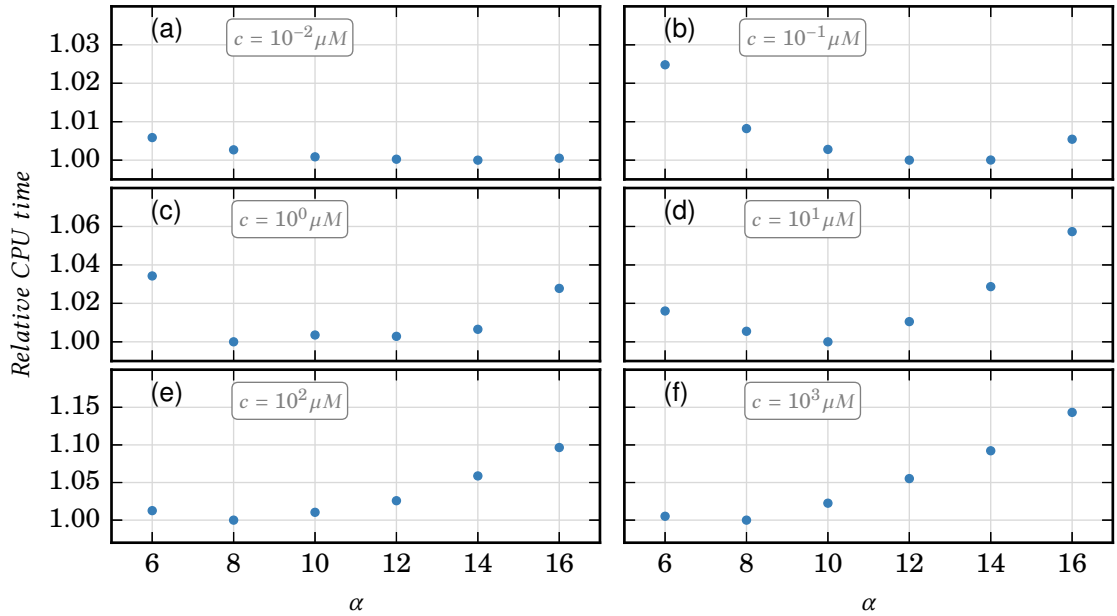} \caption{Relative CPU times required to perform the same simulation as described
in Fig. \ref{Fig:perf-1} a for different $\alpha$ values. In each
plot, the CPU times are relative to the minimum. The value $\alpha=6$
permits the construction of very small domains, even when direct time-step
integration would be preferable. The optimal value $\alpha=8.4$ found
in Fig. \ref{Fig.GF-BM} would represent the optimal value in case
the constructed domains do not burst. As $\alpha$ is increased from
its optimal value $\alpha\approx9$ the algorithm's performance decreases. }
\label{Fig:alphaPerf} 
\end{figure}

Fig. \ref{Fig:alphaPerf} shows that the optimal value of $\alpha$
lies in the range $8<\alpha<12$, in agreement with Fig. \ref{Fig.GF-BM}.
However, in the system studied here, the effect of varying $\alpha$
in $[8,12]$ on CPU performance is lower than $5\%$, and essentially
any value in this interval can be chosen. $\alpha=9$ was chosen in
the simulations shown in Fig. \ref{Fig:perf-1}. 

\newpage{}

\bibliographystyle{aip}
\bibliography{all}

\end{document}